\RequirePackage{fix-cm}
\documentclass[smallextended]{svjour3}       
\smartqed  
\usepackage{graphicx}
\usepackage{mathptmx}      
\usepackage{hyperref}       
\usepackage{url}            
\usepackage{booktabs}       
\usepackage{amsfonts}       
\usepackage{nicefrac}       
\usepackage{microtype}      

\usepackage{amssymb,amsmath}
\usepackage{graphicx}
\usepackage{subcaption}
\usepackage{wrapfig}
\graphicspath{{fig/}}

\usepackage{tikz}
\usetikzlibrary{graphs}

\usepackage{enumitem}
\setlist{leftmargin=*}

\newcommand{\PP}{{\mathcal{P}}}
\newcommand{\G}{{\mathbb G}}
\newcommand{\SO}{{\mathbb S}}
\newcommand{\TO}{{\mathbb T}}

\journalname{Journal of Healthcare Informatics Research}

\begin{document}

\title{PolSIRD: Modeling Epidemic Spread under Intervention Policies
}
\subtitle{Analyzing the First Wave of COVID-19 in the United States}


\author{Nitin Kamra \and
        Yizhou Zhang \and
        Sirisha Rambhatla \and
        Chuizheng Meng \and
        Yan Liu
}



\institute{Nitin Kamra, Yizhou Zhang, Sirisha Rambhatla, Chuizheng Meng and Yan Liu \at
              Department of Computer Science \\
              University of Southern California \\
              \email{\{nkamra,zhangyiz,sirishar,chuizhem,yanliu.cs\}@usc.edu}
}

\date{Received: date / Accepted: date}

\maketitle


\begin{abstract}
Epidemic spread in a population is traditionally modeled via compartmentalized models which represent the free evolution of disease in absence of any intervention policies. In addition, these models assume full observability of disease cases and do not account for under-reporting. We present a mathematical model, namely PolSIRD, which accounts for the under-reporting by introducing an observation mechanism. It also captures the effects of intervention policies on the disease spread parameters by leveraging intervention policy data along with the reported disease cases. Furthermore, we allow our recurrent model to learn the initial hidden state of all compartments end-to-end along with other parameters via gradient-based training. We apply our model to the spread of the recent global outbreak of COVID-19 in the United States, where our model outperforms the methods employed by the CDC in predicting the spread. We also provide counterfactual simulations from our model to analyze the effect of lifting the intervention policies prematurely and our model correctly predicts the second wave of the epidemic.
\keywords{Machine learning for epidemic spread modeling \and Epidemic spread modeling \and Spatiotemporal spread modeling \and COVID-19 \and Intervention policies for epidemics}
\end{abstract}


\section{Introduction}
\label{sec:intro}

Accurate modeling of spatiotemporal spread is critical when faced with an epidemic at a global scale such as the recent outbreak of COVID-19. This disease is caused by the SARS-CoV-2 virus, whose outbreak was first witnessed in the city of Wuhan in the Hubei province of China in December 2019. Since then the virus has spread rapidly to other parts of the world over a span of just a few months, thereby causing all countries to respond by imposing strict gathering restrictions, travel bans, stay-at-home orders and occasionally even curfews to counteract the virus spread.
While pandemics have often struck the world at different occasions throughout history, e.g. the SARS-CoV virus causing SARS in 2002 and the H1N1 swine flu in 2009--10, the SARS-CoV2 virus managed to strike the delicate balance between contagiousness and deadliness required to spread to the whole world in a matter of few months.
Given the fast spread of the disease and limited data collection capability, it has become crucial to advance our responding capabilities in various fields like contact tracing, medicine, healthcare, genome tracking and spread modeling.

Historically, pandemic projections were largely translated as linear graphs, e.g., during the Influenza pandemic in 1918. However, line diagrams are generally inadequate at predicting the spread of a spatiotemporal phenomenon like a pandemic and lead to inadequate measures being taken to handle the pandemic. Hence, it is important to progress beyond line graphs and counts of individuals in hospitals when analyzing epidemics. Hence in this article, we focus on systematic modeling of spatiotemporal spread of epidemics.

Epidemic spread has been traditionally modeled via compartmentalized models, e.g., the SIR model~\cite{kermack1927contribution}. This class of models maintains population counts in each compartment of interest (e.g. susceptible, infected and removed) and evolves them according to differential (or difference) equations. However these models do not account for under-reporting of confirmed cases or deviations from natural evolution of the disease caused by various intervention policies imposed by humans to counteract the spread. Estimating the model parameters is generally done via statistical analysis which assumes free evolution of the disease in the absence of any human intervention and requires guessing the initial hidden states of all compartments~\cite{Imai2020transmissibility}. In this work, we present our PolSIRD model which leverages intervention and disease spread data to model the disease spread. We account for under-reporting of cases via an observation mechanism and allow the model to learn the initial hidden states of all compartments along with other parameters via gradient-based training. Lastly, PolSIRD accepts intervention policy inputs and learns their effects on disease spread parameters. We apply our model to the spread of COVID-19 and also provide counterfactual simulations from our model showing the correct predictions about the second wave of the epidemic when the intervention policies are lifted prematurely.


\section{Related work}
\label{sec:related}

Modeling the spatiotemporal spread of diseases has been studied in epidemiology with the longest standing models being variants of the compartmentalized SIR model proposed by Kermack and McKendrick~\cite{kermack1927contribution}. The SIR model is a continuous-time markov model useful for studying the evolution of a disease in a single homogeneously mixing population. Over time, variants of the SIR model have been proposed and analyzed to account for discrete time, latent stages of the disease (SEIR) and lack of immunity after recovering (SIS)~\cite{allen1994some}. More advances resulting from modeling the global spread of Influenza have additionally accounted for the spread of disease via transportation between multiple heterogeneously mixing populations~\cite{rvachev1985mathematical,longini1986generalized,flahault1994mathematical}. While the parameters of these models were acquired by statistical analysis of the disease spread, we learn all parameters of PolSIRD end-to-end via gradient-based training on observed data.

Other approaches use stochastic modeling approaches based on point processes (e.g. Hawkes processes) to model disease spread in multiple populations spread across geographic locations~\cite{kim2019modeling}, however these require availability of event-level data which is often not be available due to under-reporting of cases. Other point process based approaches use aggregated event data but do not model all compartments, nor the effects of finite population sizes~\cite{linderman2015scalable}. While modifications to Hawkes processes have been proposed to have a Hawkes-SIR model~\cite{rizoiu2018sir}, these only roughly approximate the desired compartments over a finite population and have essentially been used to estimate the diameter of disease spread on the underlying graph of geographic locations. The more recently proposed GLEaM model uses a Multinomial distribution to exactly model a finite population, however, unlike PolSIRD, GLEaM does not account for under-reporting and uses statistical estimates of spread parameters instead of learning them end-to-end from data. We refer the interested reader to recent epidemiological reviews~\cite{siettos2013mathematical,chowell2016mathematical,walters2018modelling} which survey the advances in this field in more detail.

More recently, the outbreak of COVID-19 has caused numerous works to appear focusing on statistically estimating the spread and death rates from Wuhan data~\cite{verity2020estimates} and on using the SEIR model (SIR augmented with latent stage of disease) to estimate the spread in Wuhan and the rest of China~\cite{xiong2020simulating,chen2020time}. Other works have used point process based models to estimate the spread locally~\cite{lorch2020spatiotemporal}. However, the latent stage of COVID-19 is not easily observable since a latent individual is asymptomatic and yet infectious within local populations or between populations via travel. Hence, COVID-19 is plagued by under-reporting of confirmed cases which can be as high as $86\%$, as was the case in Wuhan~\cite{li2020substantial}. This along with the non-availability of enough testing equipment has made COVID-19 spread across the whole world in a matter of three months. While some recent models, like the SuEIR model~\cite{zou2020epidemic}, have modified the SEIR model to account for under-reporting of cases, they do not simultaneously learn about intervention policies.

Many works focus exclusively on studying effects of a particular intervention policy like restrictions on human mobility and foreign travel~\cite{chang2020modeling,chinazzi2020effect}
and quarantine~\cite{dandekar2020quantifying} in various countries. For instance, the recent DELPHI model~\cite{li2020overview} also accounts for an intervention policy, it does not provide any model for combining multiple intervention policies.
Our work differs from these works in that it assumes the presence of multiple intervention policies and expects only a fraction of cases to be reported. PolSIRD learns the spread, policy and reporting parameters end-to-end directly from observed data. Hence, after learning our model we can estimate the contribution of individual intervention policies towards curbing the COVID-19 spread in the United States. We are also able to run counterfactual simulations and make predictions about the effects of lifting certain intervention policies.


\section{PolSIRD model}
\label{sec:PolSIRD}

\subsection{The basic discrete-time SIR model}
\label{subsec:SIR}

The SIR compartmental model~\cite{kermack1927contribution} models the evolution of a disease in a fixed-size population in continuous time using ordinary differential equations. We describe here a discrete-time variant proposed later~\cite{longini1986generalized,allen1994some}. Assuming a population size $N$, the SIR model maintains three population compartments at all times: (a) S: Number of people susceptible to the disease, (b) I: Number of people infected by the disease and actively spreading it further, and (c) R: Number of people who have been removed either due to death or due to a full recovery leading to immunization. The state transition for individuals follows from $S \rightarrow I \rightarrow R$, hence the name SIR. At any time $t$, an individual can be in exactly one compartment and the total population is always conserved,
\begin{align}
    S(t) + I(t) + R(t) = N, \quad \forall t.
\end{align}
The Markov evolution equations of the SIR model with time step $\Delta t$ are:
\begin{align}
    S(t + \Delta t) &= S(t) - \lambda \frac{S(t)}{N} I(t) \Delta t \\
    I(t + \Delta t) &= I(t) + \lambda \frac{S(t)}{N} I(t) \Delta t - \gamma I(t) \Delta t \\
    R(t + \Delta t) &= R(t) + \gamma I(t) \Delta t
\end{align}
where the first equation assumes uniform infectious contact between individuals in compartments S and I and the last equation assumes uniform removal (via recovery or death). The parameter $\lambda \geq 0$ governs the infection rate and the parameter $\gamma \in [0,1]$ is the removal rate.
Since epidemic reported cases are generally recorded on a day-to-day basis, we will use $\Delta t = 1$ without loss of generality.

Under the SIR model, the average length of a person's infection is given by $1/\gamma$. Another important quantity is the basic reproduction number of the disease given by $r_0 = \lambda / \gamma$, i.e. the average number of people a single infectious person will infect over the course of their infection. The number $r_0$ decides if the disease will persist in the population or die out. In general, $r_0 < 1$ implies the disease automatically getting extinct, $r_0 = 1$ makes the disease endemic thereby stably persisting in the population and $r_0 > 1$ makes the disease grow exponentially and become an epidemic.

\subsection{The PolSIRD model}
\label{subsec:PolSIRD}

\begin{figure}
    \centering
    \begin{tikzpicture}
        \begin{scope}[every node/.style={circle,thick,draw, minimum size=1cm}]
            \node (S)[fill=blue!40] at (-0.5,1.5) {$S$};
            \node (Iunrep)[fill=red!50] at (3,1.5) {$I$};
            \node (Irep)[fill=red!20] at (3,-1.5) {$\tilde{I}$};
            \node (R)[fill=green!50] at (7,1.5) {$R$};
            \node (D)[fill=black!10] at (7,-1.5) {$\tilde{D}$};
            \node (Conf)[dashed] at (1,-1.5) {$\tilde{C}$};
            \node (Pol)[shape=rectangle] at (-0.5,-0.5) {$Pol$};
        \end{scope}
        
        \begin{scope}[every node/.style={circle}]
            \node (lambda)[fill=white] at (1.5,1.5) {$\{\lambda, \tilde{\lambda}\}$};
            \node (betaI)[fill=white] at (3,0) {$\beta_I$};
            \node (gamma1)[fill=white] at (5,1.5) {$\gamma$};
            \node (gamma2)[fill=white] at (5,0) {$\gamma$};
            \node (ggcbd)[fill=white] at (5,-1.5) {$\gamma (1 - \gamma_C) \beta_D$};
            \node (wb)[fill=white] at (0.5,0.5) {$w,b$};
        \end{scope}
        
        \begin{scope}[every edge/.style={draw=black,very thick}]
            \graph
            {
                (S) -- (lambda) -> (Iunrep),
                (Iunrep) -- (betaI) -> (Irep),
                (Iunrep) -- (gamma1) -> (R),
                (Irep) -- (gamma2) -> (R),
                (Irep) -- (ggcbd) -> (D),
                (betaI) ->[dashed, bend right] (Conf),
                (Pol) -- (wb) -> (lambda)
            };
        \end{scope}
    \end{tikzpicture}
    \caption{Visualization of the PolSIRD model}
    \label{fig:PolSIRD}
\end{figure}
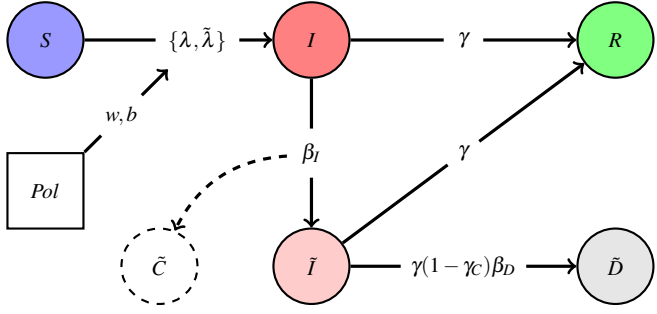

Our PolSIRD\footnote{``Pol'' stands for Policies} model augments the standard SIR model in two keys ways: (a) Designing an observation mechanism to account for under-reporting and adding two new compartments to support it, and (b) Modeling effects of intervention policies.
\\
\textbf{Observation mechanism}:
Epidemic spread data is generally plagued by under-reporting due to lack of reliable testing technology and/or due to availability of sufficient testing kits. While traditional SIR models do not account for under-reporting, we assume that the confirmed counts are only a fraction of the true total cases. We keep the susceptible (S), unreported but actively infectious (I) and the removed (R) compartments. Apart from these, we add two new compartments to account, namely the reported infectious cases $\tilde{I}$ and the reported deaths $\tilde{D}$. However, while infected individuals can recover or die and leave the $\tilde{I}$ compartment, the actual reported numbers do not track the active infectious people but rather the total number of individuals ever confirmed to be infectious till date. Hence, we also add an observable $\tilde{C}$ to keep track of the total number of confirmed cases. Note that we call $\tilde{C}$ an observable and not a compartment since individuals in it belong to $\tilde{I}, \tilde{D}$ and $R$ compartments and it double counts them for observation purposes. 
\\
\textbf{Temporal evolution}:
The population conservation can now be written as:
\begin{align}
    S(t) + I(t) + \tilde{I}(t) + R(t) + \tilde{D}(t) = N, \quad \forall t \label{eq:pop_consrv}.
\end{align}
Note that the confirmed cases observable $(\tilde{C})$ is not part of the population conservation equation since it logs individuals already present in other compartments. The Markov evolution equations of the system with state $(S,I,\tilde{I},R,\tilde{D},\tilde{C})$ for time step $=1$ day are given as:
\begin{align}
    S(t+1) &= S(t) - (\lambda I(t) + \tilde{\lambda} \tilde{I}(t)) \frac{S(t)}{N} \label{eq:suscept} \\
    I(t+1) &= I(t) + (\lambda I(t) + \tilde{\lambda} \tilde{I}(t)) \frac{S(t)}{N} - \gamma I(t) - \beta_I I(t) \label{eq:infect} \\
    \tilde{I}(t+1) &= \tilde{I}(t) - \gamma \tilde{I}(t) + \beta_I I(t) \label{eq:repinfect} \\
    R(t+1) &= R(t) + \gamma (I(t) + \tilde{I}(t)) - \gamma (1 - \gamma_C) \beta_D \tilde{I}(t) \label{eq:remove} \\
    \tilde{D}(t+1) &= \tilde{D}(t) + \gamma (1 - \gamma_C) \beta_D \tilde{I}(t) \label{eq:death} \\
    \tilde{C}(t+1) &= \tilde{C}(t) + \beta_I I(t) \label{eq:confirmed}
\end{align}
which are also described graphically in figure \ref{fig:PolSIRD}. As eq~\ref{eq:suscept} shows, we now model infections to susceptible people from both $I$ and $\tilde{I}$ compartments with their individual infection rates $\lambda, \tilde{\lambda} \geq 0$ respectively. While susceptible people go first into the unreported infectious compartment $I$, they can eventually get tested and go to the reported infectious compartment $\tilde{I}$ with a testing fraction $\beta_I \in [0,1]$ (eqs~\ref{eq:infect},\ref{eq:repinfect}). Note that the $I$ compartment also contains people who have been exposed and are infectious but not yet symptomatic, e.g., as found for the latent stage of COVID-19. Individuals in the infected compartments $I$ and $\tilde{I}$ are removed (by recovery or death) into the $R$ compartment with a rate $\gamma \in [0,1]$ (eqs~\ref{eq:remove}). However, a fraction of the deaths from confirmed infected cases $\tilde{I}$ are also reported in the $\tilde{D}$ compartment. Assuming a cure rate of $\gamma_C \in [0,1]$ and a death reporting rate of $\beta_D \in [0,1]$, the number of reported deaths per day are given as $\gamma (1 - \gamma_C) \beta_D \tilde{I}(t)$ (eq~\ref{eq:death}). Note that while we have both $\gamma_C$ and $\beta_D$ as learnable parameters, one needs to observe infections, recoveries and deaths to be able to learn the parameters separately. However, since only the confirmed infections $\tilde{C}$ and confirmed deaths $\tilde{D}$ are being reported reliably for COVID-19, one cannot disambiguate the effects of $\gamma_C$ and $\beta_D$ and our model learns the product $(1 - \gamma_C) \beta_D$ jointly as a parameter.
\\
\textbf{Influence of intervention policies}:
While most existing models simulate free evolution of a disease, the data collected generally comes from an initial freely evolving disease followed by a subsequent period of controlled spread due to execution of intervention policies by humans.
This requires us to model the influence of intervention policies, e.g. executing stay-at-home orders or closing gyms and movie theaters. Since the most rapid spread is caused by community spread, most policies are targeted towards curbing the infection rates $\lambda$ and $\tilde{\lambda}$ within a population.
Denoting the set of all policies under consideration by $\PP$, we represent the time at which policy $p \in \PP$ was enacted as $\tau_p$. In our model, a single policy $p$ is represented by four learnable parameters: $w_p, \tilde{w}_p, b_p, \tilde{b}_p$ where $w_p (\tilde{w}_p)$ represents the decay rate and $b_p (\tilde{b}_p)$ represents the steady-state reduction for $\lambda (\tilde{\lambda})$ due to the policy $p$.
While different policies can interact with each other in complex ways, we approximate the effect of multiple policies with a first-order approximation which assumes that all policies independently reduce the spread rates and multiplies the reduction to $\lambda$ (or $\tilde{\lambda}$) from all the enacted policies. Given that all policies under consideration are generally enacted within a short time frame, as has been the case with COVID-19 in the US, it is hard to disentangle the individual effects of policies under any model which considers combining multiple policies. This becomes even harder if more complex forms of interactions are modeled between policies, hence we leave the modeling of more complex policy interactions for future work and stick to our simple and more interpretable model in this work. With multiple policies being enacted at a node, both $\lambda$ and $\tilde{\lambda}$ become decaying functions of time, given by:
\begin{align}
    \lambda(t) = \lambda \prod_{p \in \PP} \left( b_p + \frac{1-b_p}{1 + e^{sgn(p) w_p (t - \tau_p)}} \right) \label{eq:pol_lmbda} \\
    \tilde{\lambda}(t) = \tilde{\lambda} \prod_{p \in \PP} \left( \tilde{b}_p + \frac{1-\tilde{b}_p}{1 + e^{sgn(p) \tilde{w}_p (t - \tau_p)}} \right) \label{eq:pol_tilde_lmbda},
\end{align}
where $sgn(p)$ is $+1$ if the policy $p$ is being enacted and $-1$ if a pre-enacted policy $p$ is being lifted.

Note that the above form ensures that enacting any policy always reduces the value of $\lambda$ and $\tilde{\lambda}$ vice-versa when lifting policies, but more importantly in a specific way. We describe the effect of eq~\ref{eq:pol_lmbda} for a single policy on $\lambda$ and the generalization with multiple policies automatically follows:
\begin{itemize}
    \item On being enacted, an intervention policy generally causes some immediate spread reduction due to a fraction of the population adopting the policy immediately. In our model, the coefficient of $\lambda$ at time $t=-\infty$ is $1$, thereby causing no reduction to the natural rate. Since the decision to enact a policy is generally rapidly taken in a few days, $\tau_p$ is not known much ahead of time before the policy is about to be enacted. Hence, we keep $t$ at $-\infty$ till $\tau_p$. When the policy is enacted at time $\tau_p$, plugging in $t=\tau_p$ immediately drops the coefficient of $\lambda$ to $\frac{1+b_p}{2}$.
    \item However, a policy continues to further reduce the spread slowly over time as the remaining population adopts the policy. Our model causes such a slow decay after time $\tau_p$ and asymptotically decays the coefficient of $\lambda$ to its steady state value of $b_p$ with an exponential rate given by $w_p$.
    \item Lastly, one can perform counterfactual simulation by lifting policy $p$ after $\lambda$ has settled to its steady state value after $p$ being enacted. This follows a similar trend of a certain fraction of people adopting the lifting immediately while other more cautious people slowly adopt it after observing the effects of lifting. On lifting policy $p$, our model increases the coefficient of $\lambda$ suddenly from $b_p$ to $\frac{1+b_p}{2}$ and then asymptotically increases it back to $1$ as time passes.
\end{itemize}
We use $\lambda(t)$ and $\tilde{\lambda}(t)$ from eqs~\ref{eq:pol_lmbda},\ref{eq:pol_tilde_lmbda} in the temporal evolution eqs~\ref{eq:suscept}--\ref{eq:confirmed} while unrolling over time.
\\
\textbf{Initialization of PolSIRD}:
Before unrolling the recurrent PolSIRD model over time, one needs to guess an initial estimate for the number of people in each compartment. This is a particularly challenging task since the true total number of people in any compartment are never observed at any time. While existing approaches~\cite{bertozzi2020challenges,zou2020epidemic} have chosen suitable initializations via approximate searches and hyperparameter tuning, we take a different route and take the initial values in each compartment at $t=0$ as learnable parameters, i.e., our model learns the fractions $s_0, i_0, \tilde{i}_0, r_0 \text{ and } \tilde{d}_0$ such that they sum upto $1$. While unrolling the model from $t=0$, we set the initial state of the model with total population $N$ as:
\begin{align}
    \{S,I,\tilde{I},R,\tilde{D},\tilde{C}\}(0) = \left\{N s_0, N i_0, N \tilde{i}_0, N r_0, N \tilde{d}_0, N \left( \tilde{i}_0 + \frac{\tilde{d}_0}{(1 - \gamma_C) \beta_D} \right) \right\}, \label{eq:init}
\end{align}
where the initialization for $\tilde{C}(0)$ is given by adding all reported active infections ($N \tilde{i}_0$) to all the reported cases which left the compartment $\tilde{I}$ to go to $\tilde{D}$ or to $R$. Since we know that people leave from $\tilde{I}$ at a rate $\gamma$, a fraction of which enters $\tilde{D}$ with rate $\gamma (1 - \gamma_C) \beta_D$, having the total reported deaths at time $t=0$ as $N \tilde{d}_0$ implies that the total number of people who ever left $\tilde{I} = \frac{N \tilde{d}_0}{(1 - \gamma_C) \beta_D}$.

\subsection{Modeling transport between heterogeneous populations}
\label{subsec:transport}

The PolSIRD model in its current form models the evolution of an epidemic at a single location where the population can be approximated to be mixing homogeneously. While previous literature~\cite{rvachev1985mathematical} also has formulations for compartmental models which can model transfer of disease between locations with heterogeneous populations, in our early experiments we observed that such advanced modeling is not required to model the spread of COVID-19 within the United States. This is primarily due to two reasons:
\begin{enumerate}
    \item First, modeling transportation of infected people within heterogeneous populations only plays a role in the early stages of the disease when seeds are being introduced into a new population. Thereafter, the evolution of the disease is heavily dominated by community spread. As will be detailed in section~\ref{subsec:training}, by the time US started recording somewhat accurate counts of infected cases and deaths in all the states (in March 2020), the disease had spread sufficiently so that the initial seed values were inconsequential and the spread was completely dominated by community spread within each homogeneous population.
    \item Secondly, transport modeling is only helpful in the absence of intervention policies which restrict travel during epidemics. However, due to international travel bans, fewer national flights and the restrictive admission to people with COVID-19 symptoms, traveling in the US during the first wave of COVID-19 was considerably lowered, thereby further deteriorating the disease spread between heterogeneous communities.
\end{enumerate}
These were also reflected in our early experiments for modeling transport between heterogeneous communities (see results in section \ref{subsec:training}). Since modeling transport only further complicated the model but did not provide any substantial gains, we have chosen to present a smaller but more interpretable PolSIRD model in this work. The more rigorous GraphPolSIRD model which also considers transportation between different states has been described in the appendix for the interested reader.

\subsection{Training}
\label{subsec:learning}

The PolSIRD model takes as input the time-steps at which each policy was enacted and unrolls the temporal evolution model defined by eqns~\ref{eq:suscept}--\ref{eq:pol_tilde_lmbda} for $T$ time steps. The model is trained to learn all its parameters $\theta = \{ \lambda, \tilde{\lambda}, \gamma, \beta_I, \beta_D, \gamma_C \} \cup \{ w_p, b_p, \tilde{w}_p, \tilde{b}_p \}_{p \in \PP} \cup \{ s_0, i_0, \tilde{i}_0, r_0, \tilde{d}_0 \}$ by minimizing the Mean Absolute Error Ratio (MAER) on the number of confirmed cases $\tilde{C}(t)$ and the number of reported deaths $\tilde{D}(t)$ at all time steps:
\begin{align}
    \min_{\theta} \sum_{t=1}^T \left( \frac{| \hat{C}(t) - \tilde{C}(t) |}{\max(\hat{C}(t), 1)} + \frac{| \hat{D}(t) - \tilde{D}(t) |}{\max(\hat{D}(t), 1)} \right), \label{eq:MAER}
\end{align}
where the $\max(\cdot)$ term in the denominator serves to handle nodes with $0$ reported cases and a $hat(\hat{})$ on the top denotes the observed ground truth value. Note that while other works~\cite{zou2020epidemic} have used the Mean Squared Error (MSE) to train similar models, we found in our preliminary training runs that using MSE forces the model to become very sensitive to noise and outliers in the data.
While one can train the model independently at any location, in practice we train models jointly across multiple locations (e.g. on all US states). While each location maintains its own copy of the temporal evolution parameters and the initial state parameters for compartments since these can vary significantly across locations, we enforce that policies have similar effects at all locations and leverage additional data for learning about policy effects by sharing the same copy of policy decay rates and steady-state reductions amongst all locations for each policy\footnote{Note that all policies still have their own unique decay rates and steady-state reductions. We do not share those parameters amongst different policies.}. Further, this makes the use of MAER loss more suitable over alternatives like MSE or MAE (Mean Absolute Error) since the denominator in MAER normalizes the loss values from each location to be between 0 and 1, thereby obviating the need to specify scaling coefficients between losses from different locations.


\section{Experiments}
\label{sec:exp}

\subsection{Datasets}
\label{subsec:datasets}

We apply our PolSIRD model to predicting the first wave of COVID-19 and the corresponding spread parameters for all states in the United States. We use the following datasets for training our model:
\begin{enumerate}
    \item \textbf{COVID-19 confirmed cases}: We use confirmed case counts and death counts for the United States maintained by the CSSE at John Hopkins University~\cite{dong2020interactive}.
    \item \textbf{US Population data}: We use the latest estimates from the United States Census Bureau~\cite{bureau2020population}.
    \item \textbf{US Intervention policy data}: We use data aggregated from various sources by researchers at John Hopkins University~\cite{killeen2020countylevel}. The policies we consider are: (a) stay-at-home, (b) ban $>50$ gatherings, (c) ban $>500$ gatherings, (d) public school closure, (e) restaurant dine-in closure, and (f) entertainment/gym closure.
    \item \textbf{PlaceIQ movement data}: We use the exposure indices derived from PlaceIQ movement data which describe travel within the United States on the basis of smartphone movements. These are derived from anonymized, aggregated smartphone movement data provided by PlaceIQ~\cite{placeiq2020USmobility} and used only in our more rigorous GraphPolSIRD model mentioned in section~\ref{subsec:transport} and described in the appendix.
\end{enumerate}

\subsection{Training results}
\label{subsec:training}

\begin{figure*}[th]
    \centering
    \begin{subfigure}[h]{.32\textwidth}
        \centering
        \includegraphics[width=\linewidth]{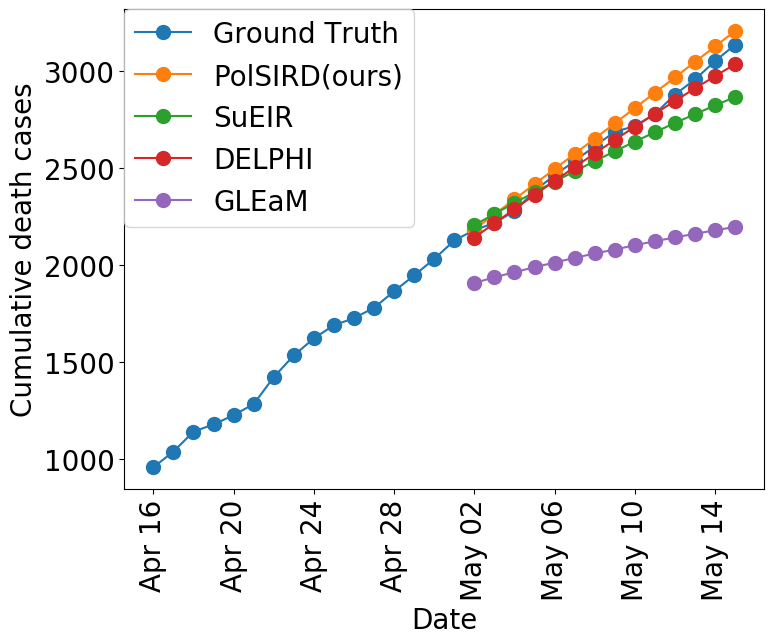}
        \caption{}
        \label{fig:deaths_CA}
    \end{subfigure}
    \begin{subfigure}[h]{.32\textwidth}
        \centering
        \includegraphics[width=\linewidth]{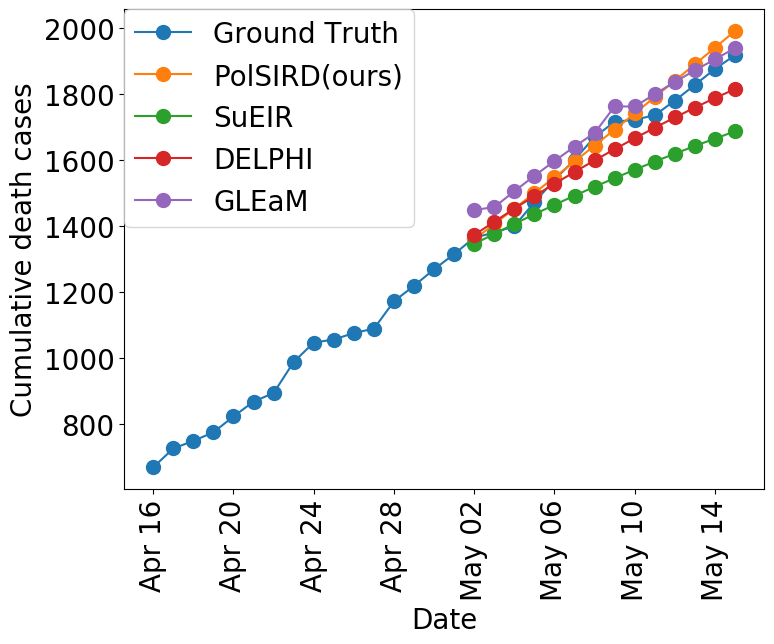}
        \caption{}
        \label{fig:deaths_FL}
    \end{subfigure}
    \begin{subfigure}[h]{.32\textwidth}
        \centering
        \includegraphics[width=\linewidth]{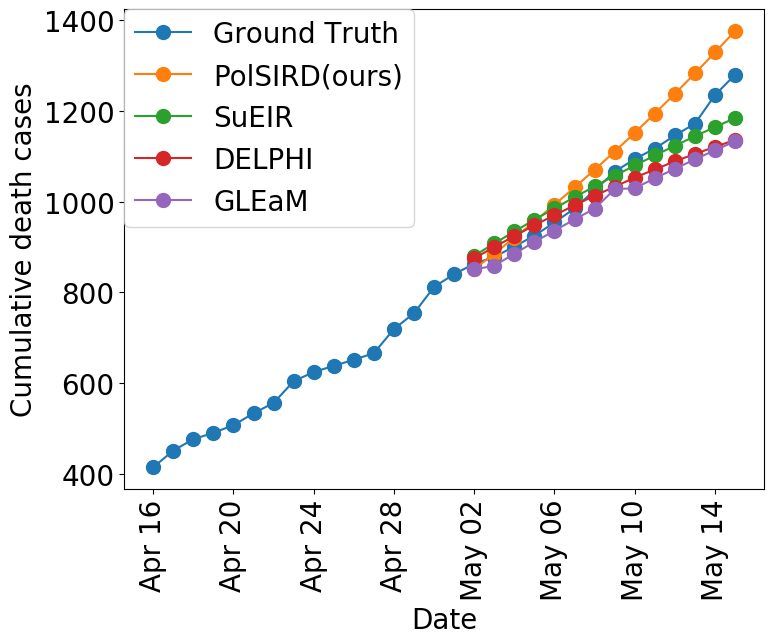}
        \caption{}
        \label{fig:deaths_TX}
    \end{subfigure}
    \caption{Death predictions from all models on: (a) California, (b) Florida, and (c) Texas}
    \label{fig:deaths}
\end{figure*}

\textbf{Implementation details}: We trained our PolSIRD model as described in the previous section to learn the spread, reporting and policy parameters for COVID-19. Since reporting in the US states before mid of March 2020 was very inaccurate and sparse, and after May 20th several states started lifting many intervention policies, we extracted the COVID-19 confirmed cases from March 22nd to May 20th and split them as follows: Train data (March 22--May 1) and Test data (May 2--May 20). Further, preliminary training runs showed a considerable variance in learnt parameters when using very long sequences from the reported COVID-19 data indicating high non-stationarity in the underlying temporal process. Hence, for reporting final results we trained our model on a recent sub-sequence extracted from April 22 onwards from the train split. We minimized the MAER as defined in eq~\ref{eq:MAER} to learn all trainable parameters in an end-to-end fashion for $16,000$ epochs with the Adam optimizer having a learning rate of $0.003$.
We obtained a small final test MAER of 0.0825 (i.e. an 8\% relative prediction error) jointly on the confirmed and death cases, thereby showing that PolSIRD is a suitable model for predicting epidemic spread.
\\
\textbf{Comparing with baselines}: We compare our model specifically against several baselines generated by other institutes and which were being used to inform the decisions made by the Centers for Disease Control and Prevention (CDC): (a) SuEIR model proposed by Zou et al.~\cite{zou2020epidemic}, (b) GLEaM model proposed by Balcan et al.~\cite{balcan2010modeling} and, (c) DELPHI model proposed by Li~\cite{li2020overview}. Since the CDC only maintains predictions of future deaths, we directly used the predicted death cases from all the baselines being used by the CDC (taken from \cite{covid19hub}) and compare them to our model's predictions. For this we only used the predictions made from May 2nd onwards and ensure that we used versions of all models which had only been trained on the time-series uptil May 1st. We show the MAER and the Root Mean Square Error (RMSE) for all models in Table~\ref{tab:MAER} averaged across all the US states. We also additionally show both metrics averaged across the best 40 states for each model. This removes certain states which have abnormalties in reporting or very few reported deaths causing the models to have a high error on those states. Our PolSIRD model clearly outperforms all the other state-of-the-art baselines in terms of both metrics on all 51 states (counting the District of Columbia as a state) and also on the top-40 states. We show the death predictions from all models on several states in figure~\ref{fig:deaths}. Lastly, we also present the results from the GraphPolSIRD model mentioned in section~\ref{subsec:transport}, which takes traveling of people between states into account, in Table~\ref{tab:MAER}. We achieve very similar results from this model as the PolSIRD model with minor differences in RMSE values. Since modeling transport only further complicated the model but did not provide any substantial gains in prediction performance, we have chosen to present the smaller but more interpretable PolSIRD model in this work. However, the GraphPolSIRD model is presented in detail in the appendix for the interested reader.

\begin{table}[th]
    \centering
    \caption{The MAER and RMSE metrics of PolSIRD compared to other state-of-the-art baselines being used by the CDC to inform their decisions.}
    \label{tab:MAER}
    \begin{tabular}{|c|c|c|c|c|}
        \hline
        Method & MAER (all states) & RMSE (all states) & MAER (top-40) & RMSE (top-40) \\
        \hline
        PolSIRD (ours) & \textbf{0.076} & \textbf{184.147} & \textbf{0.046} & 36.393 \\
        SuEIR & 0.081 & 653.304 & 0.059 & \textbf{35.273} \\
        GLEaM & 0.223 & 1400.22 & 0.144 & 88.231 \\
        DELPHI & 0.142 & 242.815 & 0.092 & 53.41 \\
        \hline
        GraphPolSIRD (ours) & \textbf{0.076} & \textbf{186.303} & \textbf{0.046} & 36.164 \\
        \hline
    \end{tabular}
\end{table}

\subsection{Parameter estimation}
\label{subsec:param_est}

\begin{figure*}[ht!]
    \centering
    \begin{subfigure}[h]{.75\textwidth}
        \centering
        \includegraphics[width=\linewidth]{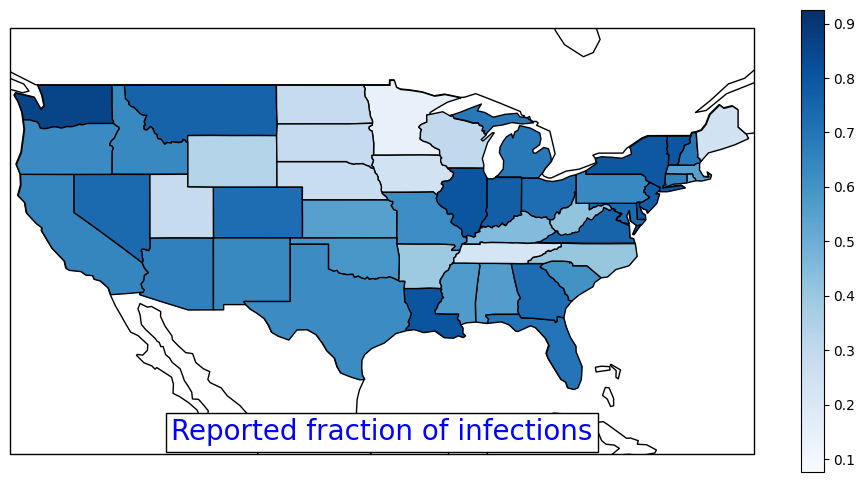}
        \caption{}
        \label{fig:viz_betaI}
    \end{subfigure}
    \begin{subfigure}[h]{.75\textwidth}
        \centering
        \includegraphics[width=\linewidth]{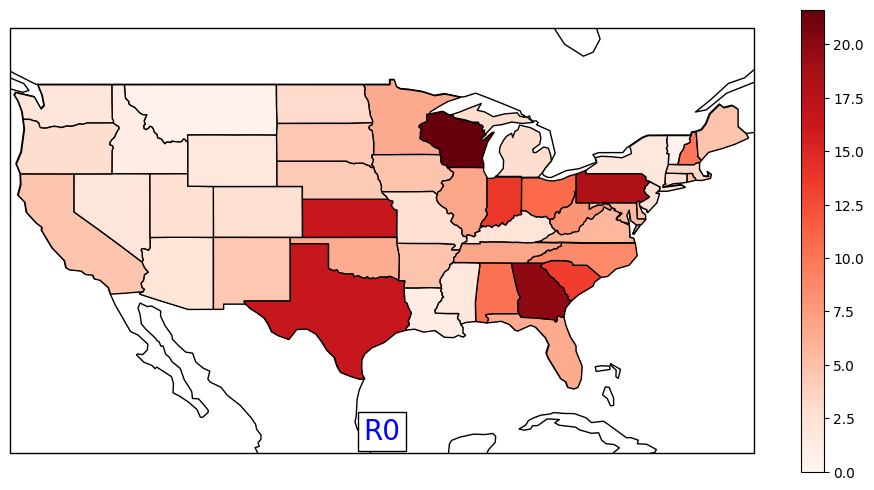}
        \caption{}
        \label{fig:viz_r0}
    \end{subfigure}
    \begin{subfigure}[h]{.75\textwidth}
        \centering
        \includegraphics[width=\linewidth]{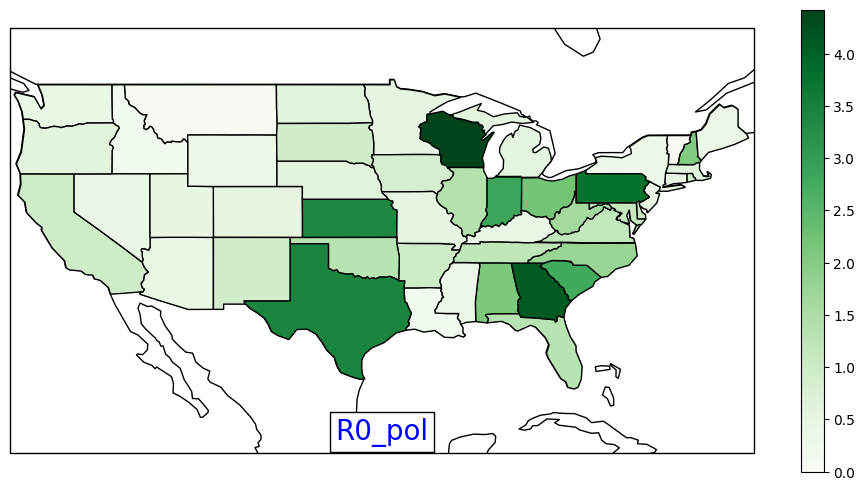}
        \caption{}
        \label{fig:viz_r0_pol}
    \end{subfigure}
    \caption{Intensity maps visualizing the estimated values of: (a) the reported fraction of confirmed cases $\beta_I$, (b) the natural reproductive number $r_0$, and (c) the reproductive number $r_0$ after intervention policies reach a near steady-state across all the US states.}
    \label{fig:viz_maps}
\end{figure*}

We report the parameter values learnt by our model averaged over the US states in Table~\ref{tab:avg_params}. 
Our estimated mean disease lifetime given by $1/\gamma = 17.86$ days is a lies in the statistical range of $17.8$ to $24.7$ days estimated by \cite{verity2020estimates}. Based on $\beta_I$, our model estimates about $61\%$ reported confirmed cases in the US, which is fortunately a significant fraction of the true confirmed cases. Further, we calculate the natural reproductive number ($r_0$) of COVID-19 in the US. While this quantity representing the total number of secondary infections by a single infected individual is given by $\lambda/\gamma$ for the standard SIR model, for the PolSIRD model this quantity is given by: $\frac{\lambda}{\gamma+\beta_I} + \frac{\tilde{\lambda}}{\gamma} \frac{\beta_I}{\gamma + \beta_I}$. Our estimated value averaged across all states: $r_0 = 5.899$, is about double the $2.9$ value reported for Wuhan~\cite{Imai2020transmissibility} indicating a higher spread rate if there were no intervention policies at all\footnote{However the actual reproduction number is less and is explored in section~\ref{subsec:policy}.}. Since both $\beta_I$ and $r_0$ values vary considerably across states, we visualize them across all US states in figures~\ref{fig:viz_betaI} and \ref{fig:viz_r0} respectively.
The high values of $r_0$ in New York, New Jersey, Texas, Atlanta and Wisconsin are consistent with the known high spread rates in these states. While the reporting fractions are high in Washington, Louisiana, Illinois and New York, many other states suffer from heavy under-reporting. The first major way in which our model can aid decision making is by identifying states where the reporting rate is low. In such cases, a first actionable measure for such states would be to provision more resources for testing and reporting cases in order to gauge the spread of COVID-19 more accurately.

\begin{table}[h]
    \centering
    \caption{Learnt parameter values from PolSIRD averaged over the US states.}
    \label{tab:avg_params}
    \begin{tabular}{|c|ccccccc|}
        \hline
        Parameter & $\tilde{\lambda}$ & $\lambda$ & $\gamma$ & $\gamma_C$ & $\beta_I$ & $\beta_D$ & $r_0$ \\
        \hline
        Mean across states & $0.177$ & $0.178$ & $0.056$ & $0.899$ & $0.61$ & $0.628$ & $5.899$ \\
        Stddev across states & $\pm 0.094$ & $\pm 0.13$ & $\pm 0.047$ & $\pm 0.057$ & $\pm 0.209$ & $\pm 0.134$ & $\pm 4.798$ \\
        \hline
    \end{tabular}
\end{table}

\subsection{Intervention policy analysis}
\label{subsec:policy}

\begin{figure*}[ht!]
    \centering
    \begin{subfigure}[h]{.32\textwidth}
        \centering
        \includegraphics[width=\linewidth]{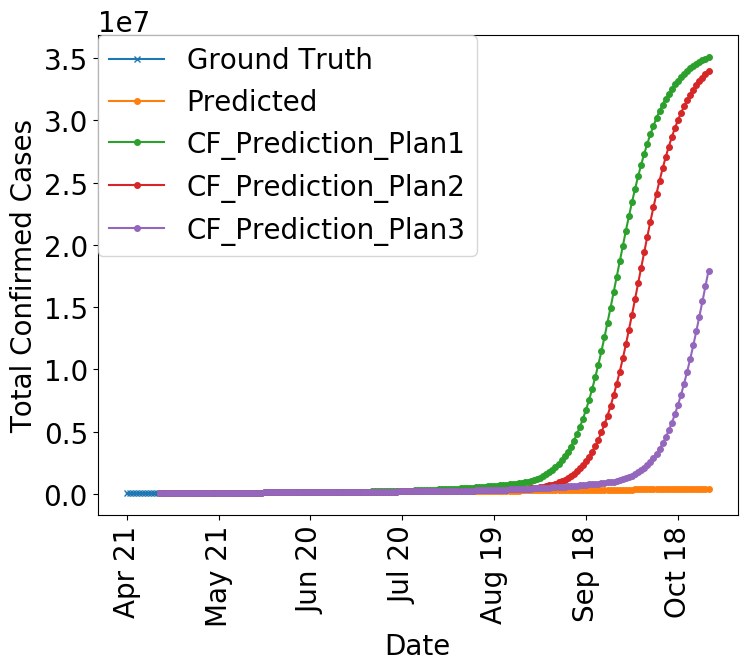}
        \caption{}
    \end{subfigure}
    \begin{subfigure}[h]{.32\textwidth}
        \centering
        \includegraphics[width=\linewidth]{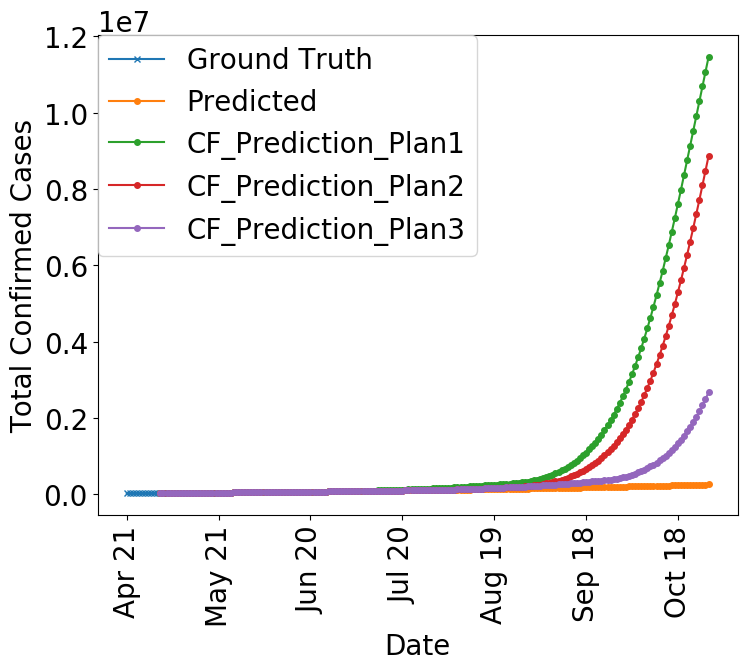}
        \caption{}
    \end{subfigure}
    \begin{subfigure}[h]{.32\textwidth}
        \centering
        \includegraphics[width=\linewidth]{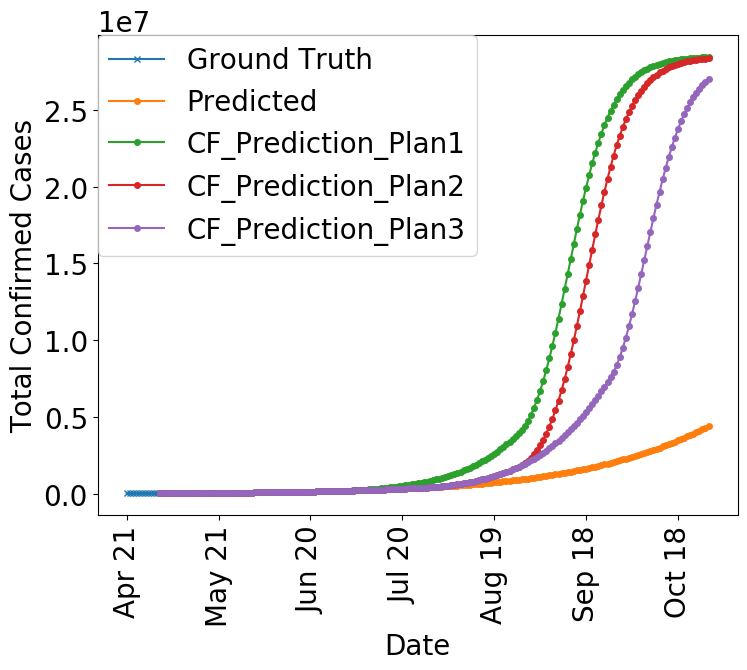}
        \caption{}
    \end{subfigure}
    \caption{Predictions on (a) California, (b) Florida and (c) Texas showing the disease progression under intervention and the ensuing exponential growth under various re-opening plans.}
    \label{fig:reopen}
\end{figure*}

\textbf{Effects of policies}: We next analyze the effects of intervention policies on the infection rates and the reproduction number. Table~\ref{tab:policy} summarizes the steady-state decays to $\lambda, \tilde{\lambda}$ from each policy. We observed from our experiments that disambiguating the contribution of various policies is hard since they were enacted between 6--20 March with very short durations between them to observe their individual effects. Hence, we focus on the overall steady-state decay due to all policies combined together in Table~\ref{tab:policy_overall}.
Overall, the policies have been quite effective leading to a reduction in spread of about $79\%$ from reported cases and $93\%$ from unreported cases. This is expected because people with confirmed infection tend to isolate themselves while asymptomatic unreported cases unknowingly spread the disease. This unreported spread gets curbed to a larger extent by intervention policies. The new reproduction number $r_0$ dropped down to $1.159$ averaged across the US states, which while much smaller than its natural value is still $>1$, indicating that the disease continued to progress, albeit at a slower rate. However several states had $r_0$ value less than 1 (see figure~\ref{fig:viz_r0_pol}), which became a motivating factor for the governments of these states to begin formulating plans to lift the policies and re-open the states.
\\
\textbf{Counterfactual re-opening of states}: Most re-opening plans devised were multi-stage and were to remove intervention policies sequentially. Around May 20th, many states were already in stages 1 or 2 of re-opening which prepped workplaces to follow social distancing guidelines and allowed opening small low-risk workplaces and businesses respectively~\cite{smith2020reopening}. To understand the effects of eventually moving into stages 3 (opening restaurants for dine-in, gyms and entertainment venues) and 4 (lifting stay-at-home order, bans on large gatherings and public school closures), we now perform counterfactual simulations from our trained model to lift the intervention policies. We execute three counterfactual stage 3 and 4 plans as follows (all stage plans are in reference to May 1, 2020):
\begin{enumerate}
    \item \textbf{Plan 1}: Enter stage 3 in 60 days and stage 4 in 120 days,
    \item \textbf{Plan 2}: Enter stage 3 in 90 days and stage 4 in 120 days, and
    \item \textbf{Plan 3}: Enter stage 3 in 90 days and stage 4 in 150 days.
\end{enumerate}
The results of our counterfactual simulations when all policies are in place and when the above plans are followed are shown on several states in figure~\ref{fig:reopen}.
While the disease progression is slowed under policy intervention, we observe that the number of infected cases begins to rise exponentially if the policies are removed following any of the above three plans; the key takeaway being that it would have been detrimental to remove the intervention policies in the near future after May. This is another use case of our model that we can make approximate counterfactual predictions for the future. To be cautious and not cause a second wave of COVID-19 infections, our model suggested deferring the stages 3 and 4 of re-opening to more than 150 days (from May 1, 2020) in at least the states which have a current $r_0 \geq 1$ (figure~\ref{fig:viz_r0_pol}). However, in retrospect, many states did enter their stages 3 and 4 shortly after May and suffered from an impending second wave of COVID-19 after which many intervention policies were restored back all over the United States.

\begin{table}[h]
    \centering
    \caption{Steady state policy coefficients for $\lambda$ and $\tilde{\lambda}$.}
    \label{tab:policy}
    \begin{tabular}{|c|cccccc|}
        \hline
        Policy $p$ & stay-at & ban $>50$ & ban $>500$ & public school & restaurant & entertainment \\
         & -home & gatherings & gatherings & closure & dine-in closure & /gym closure \\
        \hline
        $b_p$ & 0.559 & 0.738 & 0.708 & 0.52 & 0.667 & 0.685 \\
        $\tilde{b}_p$ & 0.815 & 0.812 & 0.817 & 0.71 & 0.737 & 0.738 \\
        \hline
    \end{tabular}
    \caption{Overall reduction in spread rate and reproduction number.}
    \label{tab:policy_overall}
    \begin{tabular}{|c|ccc|}
        \hline
         & $\tilde{\lambda}$ & $\lambda$ & $r_0$ \\
        \hline
        Before policies & $0.177 \pm 0.094$ & $0.178 \pm 0.13$ & $5.899 \pm 4.798$ \\
        After policies & $0.037 \pm 0.019$ & $0.012 \pm 0.009$ & $1.159 \pm 1.016$ \\
        \hline
    \end{tabular}
\end{table}

\section{Conclusion and Broader Impact}
\label{sec:conc}

In this paper, we have presented a mathematical model, PolSIRD, which models the spread of epidemics within a population. We model the spread of the disease under intervention policies and under-reporting while trying to analyze the plans being formulated to re-open the US states after the first wave of COVID-19. Our method successfully outperforms the methods being used by the Center for Disease Control and Prevention for making spread predictions on COVID-19 and provides counterfactual simulations for understanding the effects of lifting the intervention policies.
We believe that both the state governments and the decision making authorities (e.g., Centers for Disease Control and Prevention) can potentially benefit from our modeling approach in other similar future outbreaks. While we cannot guarantee that all the parameters we have estimated from our model can be deemed precisely accurate, we have made a significant effort to verify that they are in agreement with parameters being estimated from other statistical studies. However, they may still be affected by: (a) biases in data reporting procedures, (b) large variations in availability of testing equipment across the US states, (c) interactions between various policies and, (d) the short time-window within which all policies were enacted in all the US states. But lacking a more elaborate data collection procedure, we defer the study of more advanced models which work around the above mentioned issues to future work.


%
\section*{Conflict of interest}

The authors declare that they have no conflict of interest.

\bibliography{references}   
\bibliographystyle{ieeetr}

\appendix

\section{Appendix}

\subsection{GraphPolSIRD: Spatial transport with PolSIRD}
\label{subsec:graphpolsird}

In this section, we describe the more detailed GraphPolSIRD model which collectively models both the temporal and the spatial evolution of a pandemic. Consider a set of $M$ heterogeneous populations (e.g., the $M=50$ for the US states) as nodes in a graph $\G$. We assume that the population on a node can be assumed to mix homogeneously within itself while the populations between two nodes require traveling to mix. There exists a directed edge from node $i$ to node $j$ in the graph if people can travel from $i$ to $j$. The edge weight denoted by $f_{ij}$ is proportional to the frequency at which residents of node $i$ travel to node $j$ and come back. Note that $f_{ij}$ may not equal $f_{ji}$ since some states like New York may be popular tourist/business destinations while others like Alaska might experience limited number of visitors (hence $f_{AK,NY} \gg f_{NY,AK}$). An estimate of these travel frequencies is required from travel data for the GraphPolSIRD model. For our experiment results in section~\ref{subsec:training}, we used the PlaceIQ movement data derived from anonymized, aggregated smartphone movements provide by PlaceIQ~\cite{placeiq2020USmobility}. Note that while we assume these travel frequencies $f_{ij}$ to be averaged constants independent of time, in practice this may not be the case. If non-aggregated daily travel frequencies are available, our GraphPolSIRD model also admits using them instead. However, for our application to Covid-19 modeling, only aggregated travel frequencies were available and hence we describe a simpler version of GraphPolSIRD with constant graph edge weights $f_{ij}$ independent of time.

At any time $t$, the state $G(t)$ of graph $\G$ is described by the compartment populations at all the nodes i.e. $G(t) = \{S_i(t), I_i(t), \tilde{I}_i(t), R_i(t), \tilde{D}_i(t), \tilde{C}_i(t)\}_{i \in [M]}$. The population conservation equation still applies independently at all nodes:
\begin{align}
    S_i(t) + I_i(t) + \tilde{I}_i(t) + R_i(t) + \tilde{D}_i(t) = N_i, \quad \forall i \in [M], \forall t.
\end{align}

The evolution from time $t$ to $t+1$ is now governed by a cascade of two operators, denoted as $\TO$ (the temporal evolution operator) and $\SO$ (the spatial evolution operator), i.e.,
\begin{align}
    G(t+1) = \TO(\SO(G(t))), \forall t.
\end{align}
The temporal operator $\TO$ applies to each node individually and is governed by the same set of equations \ref{eq:suscept}--\ref{eq:pol_tilde_lmbda} described in section~\ref{sec:PolSIRD}. Here we describe the spatial operator $\SO$ which applies to the graph $\G$ as a whole and models the effect of epidemic spread due to travel between the heterogeneously mixing populations at different nodes.

The spatial operator $\SO$ affects only the susceptible and the unreported infectious compartments at a node due to the presence of infected population traveling to it from adjacent nodes. We assume that reported infected individuals refrain from traveling (or the number of such individuals traveling is small enough to be neglected), hence the inter-node spread happens primarily due to unreported infected individuals traveling between nodes. Hence, the spatial operator reduces the susceptible population and increases the infected population at each time step at every node $i \in [M]$ as follows:
\begin{align}
    S_{i,\SO}(t) &= S_i(t) - \sum_{j \in [M]\setminus\{i\}} \sigma f_{ji} \frac{I_j(t)}{N_j - \tilde{I}_j(t) - \tilde{D}_j(t)} \\
    I_{i,\SO}(t) &= I_i(t) + \sum_{j \in [M]\setminus\{i\}} \sigma f_{ji} \frac{I_j(t)}{N_j - \tilde{I}_j(t) - \tilde{D}_j(t)},
\end{align}
where the term being subtracted from $S_i(t)$ and being added to $I_i(t)$ are the number of individuals getting affected at node $i$ at time step $t$ due to unreported infected individuals traveling from neighboring nodes $j \in [M]\setminus\{i\}$. Note that the term $N_j - \tilde{I}_j(t) - \tilde{D}_j(t)$ is the total traveling population (since reported infected individuals and dead individuals do not travel) and the term $\frac{I_j(t)}{N_j - \tilde{I}_j(t) - \tilde{D}_j(t)}$ gives the number of unreported infected population as a fraction of the total population traveling. When multiplied by the average number of people travelling from node $j$ to node $i$, i.e. $f_{ji}$ (a.k.a. travel frequency), the term $f_{ji} \frac{I_j(t)}{N_j - \tilde{I}_j(t) - \tilde{D}_j(t)}$ gives the number of unreported infected people traveling from node $j$ to node $i$. The constant $\sigma \in (0, \infty)$ is an additional learnable parameter of the GraphPolSIRD model which stands for the number of people infected at the destination node per traveling person from a source node. This parameter is learnt end-to-end with gradient descent along with the other learnable parameters described in section~\ref{subsec:learning}. Note that since we have the learnable paramter $\sigma$ multiplying with the travel frequencies $f_{ji}$, it can account for any magnitude changes in the $f_{ij}$ coefficients. This implies that we do not necessarily need exact travel frequencies. Instead one can use any available quantities proportional to them. This is a very useful property since the PlaceIQ movement data~\cite{placeiq2020USmobility} does not provide us exact travel frequencies but rather the fraction of smartphones in state A on the current day that pinged in any of the other states in the last 15 days. This quantity is roughly proportional to the travel frequencies but does not reflect their exact magnitude, however our learnable parameter $\sigma$ can to adapt to any scaling in the magnitudes.

Finally, the spatial operator does not change any other compartments or observables other than $S$ and $I$. Hence, after the spatial operator $\SO$, we can apply the temporal operator $\TO$ to the graph state $G(t) = \{S_{i,\SO}(t), I_{i,\SO}(t), \tilde{I}_i(t), R_i(t), \tilde{D}_i(t), \tilde{C}_i(t)\}_{i \in [M]}$ as described in the main text to model the transition to time step $t+1$.

\end{document}